\documentclass[reprint,amsmath,amssymb,prb]{revtex4-1}
\usepackage{amsmath,amssymb,amsfonts}
\usepackage{graphicx}
\usepackage{dcolumn}
\usepackage{bm}
\usepackage{morefloats,multirow,titlesec}
\usepackage[hang,raggedright]{subfigure}
\usepackage[export]{adjustbox}
\usepackage{tabularx,ragged2e,booktabs}
\usepackage{hyperref}
\usepackage[mathlines]{lineno}

\begin{document}

\preprint{APS/123-QED}

\title{Diffuson contribution to anomalous Hall effect in disordered Co$_{2}$FeSi thin films }
\author{Binoy Krishna Hazra}
\affiliation{School of Physics, University of Hyderabad, Hyderabad-500046, India}
\author{S. N. Kaul}
\email{sn.kaul@uohyd.ac.in}
\affiliation{School of Physics, University of Hyderabad, Hyderabad-500046, India}
\author{S. Srinath}
\email{srinath@uohyd.ac.in}
\affiliation{School of Physics, University of Hyderabad, Hyderabad-500046, India}
\author{M. Manivel Raja}%
\affiliation{Defence Metallurgical Research Laboratory, Hyderabad-500058, India}
\author{R. Rawat, Archana Lakhani}
\affiliation{UGC-DAE Consortium for Scientific Research, Indore-452001, India}

\date{\today}

\begin{abstract}

A wide variation in the disorder strength, as inferred from an order of magnitude variation in the longitudinal resistivity ($\rho_{xx}$) of Co$_2$FeSi (CFS) Huesler alloy thin films of fixed (50 nm) thickness, has been achieved by growing these films on Si(111) substrates at substrate temperatures (T$_{S}$) ranging from room temperature (RT) to 600$^{\circ}$C. An in-depth study of the influence of disorder on anomalous Hall resistivity ($\rho_{xy}^{AH}$), $\rho_{xx}$ and magnetoresistance (MR), enabled by this approach, reveals the following. The side-jump mechanism gives a dominant contribution to anomalous Hall resistivity (AHR) in the CFS thin films, regardless of the degree of disorder present (i.e., whether they are in the amorphous (high-resistivity) or crystalline (low-resistivity) state). A new and novel contribution to both $\rho_{xx}$ and $\rho_{xy}^{AH}$, characterized by the logarithmic temperature ($-lnT$) dependence at temperatures below the minimum (T$<$T$_{min}$), exclusive to the amorphous CFS films, originates from the scattering of conduction electrons from the diffusive hydrodynamic modes associated with the longitudinal component of magnetization, called `diffusons'. In these amorphous CFS films, the electron-diffuson, $e-d$, scattering and weak localization (WL) mechanisms compete with that arising from the inelastic electron-magnon, $e-m$, scattering to produce the minimum in $\rho_{xx}(T)$, whereas the minimum in $\rho_{xy}^{AH}(T)$ is caused by the competing contributions from the $e-d$ and $e-m$ scattering, as WL \textit{does not make any contribution} to AHR. These results thus vindicate the long-standing, but hitherto unverified, theoretical prediction that in \textit{high-resistivity} metallic ferromagnets in which the \textit{side-jump} mechanism prevails, WL correction to $\rho_{xy}^{AH}$ \textit{vanishes} even when the WL effect contributes to $\rho_{xx}$ for T $\lesssim$ T$_{min}$. In sharp contrast, in crystalline films, enhanced electron-electron Coulomb interaction (EEI), which is basically responsible for the resistivity minimum, makes no contribution to $\rho_{xy}^{AH}(T)$ with the result that AHR does not exhibit a minimum. The conventional $\rho_{xy}^{AH}$ = $f(\rho_{xx})$ scaling breaks down completely in the present case, more so in the strongly disordered (amorphous) CFS films. Instead, when $\rho_{xy}^{AH}(T)$ is corrected for the $e-d$ contribution and $\rho_{xx}(T)$ for both $e-d$ and WL contributions (only EEI) in the amorphous (crystalline) films, and the AH coefficient, $R_{A}(T)$ = $\rho_{xy}^{AH}(T)$ / $4\pi M_{s}(T)$, (calculated from the \textit{corrected} $\rho_{xy}^{AH}$ and spontaneous magnetization, $M_{s}$), perfectly scales with $\rho_{xxT}$, the temperature-dependent part of the \textit{corrected} $\rho_{xx}$, for all the CFS thin films. 
\end{abstract}

\maketitle

\section{\label{sec:level1} Introduction}

Anomalous Hall effect (AHE) in ferromagnets is known to arise from two basic mechanisms classified as the extrinsic and intrinsic mechanisms. The skew scattering (Sk) \cite{Smit1955} and side-jump (Sj) \cite{Berger1970}, originating from the asymmetric scattering of conduction electrons from impurities caused by the spin orbit interaction (SOI), fall within the extrinsic category whereas the SOI-induced anomalous transverse velocity of the Block electrons (the so-called KL mechanism) \cite{KL1954}, recently reinterpreted as an integral of the Berry-phase curvature (Bc) over occupied electronic bands within the Brillouin zone \cite{Yao2004,Naga2010,Xiao2010,Kuebler2012}, constitutes the intrinsic mechanism. While the extrinsic Sk mechanism predicts that $R_{A}$ $\propto $ $\rho_{xx}$, extrinsic Sj mechanism, like the intrinsic KL and Bc mechanisms, is characterized by the relation $R_{A}$ $\propto $ $\rho_{xx}^{2}$ between the anomalous Hall coefficient, $R_{A}$, and longitudinal resistivity, $\rho_{xx}$. Based on these theoretical predictions, measured $R_{A}$ or the anomalous Hall resistivity, $\rho_{xy}^{AH}$, is often analyzed \cite{Naga2010,Kaul1979} in terms of the expression 

\begin{equation}
R_{A} = \rho_{xy}^{AH} / 4 \pi M_{s} = a~\rho_{xx} + b~\rho_{xx}^2 
\end{equation}  
 
where $M_{s}$ is the spontaneous magnetization and the coefficient $a$ ($b$) is a direct measure of the strength of the skew scattering (Sj \textit{extrinsic} and KL/Bc \textit{intrinsic} contributions). The main problem with the use of Eq.(1) is that it does not permit an unambiguous separation of the Sj (extrinsic) from the KL/Bc (intrinsic) contribution because these mechanisms yield the same power law dependence on $\rho_{xx}$. Recently, a solution to this problem has been sought in terms of a generalized scaling relation, $\rho_{xy}^{AH}$ = $f(\rho_{xx})$, derived from a first principles calculation of AHE \cite{Naga2010,Xiao2010,Onoda2006,Hou2015,Hyodo2016}. Such an approach proves useful \cite{Miya2007,Tian2009,Xiong2010,Vidal2011,Ye2012,Presti2014} only in pure (ordered) ferromagnetic systems in which ballistic transport governs $\rho_{xx}(T)$. Strong departures from the predicted scaling behavior have been observed particularly in disordered ferromagnets (e.g., in amorphous CoFeB \cite{Zhu2015}, Heusler alloy thin films \cite{Obaida2011,Zhu2012,Gabor2015}, Fe \cite{Lin2016} and Ni\cite{Guo2012} ultra-thin films, FePt\cite{Lu2013} thin films) that exhibit a minimum at low temperatures either in $\rho_{xx}(T)$ or in $\rho_{xy}^{AH}(T)$ or in both. Considering that the previous works on different disordered ferromagnetic systems suffer from the inextricably intertwined effects of composition, disorder and spatial dimensionality, a systematic investigation on a system of fixed composition and film thickness, prepared in different states of disorder, is thus needed for a deeper understanding of the AHE phenomenon and the role of disorder.

In a disordered metallic system, resistivity minimum at a temperature $T_{min}$ arises from a competition between the quantum corrections (QCs) such as weak localization (WL) and enhanced electron-electron Coulomb interaction (EEI), which increase $\rho_{xx}$ when the temperature is reduced below $T_{min}$, and the electron-electron, electron-phonon, electron-magnon inelastic scattering processes \cite{LeeTVR1985,Babu1999,Srini1999} that destroy phase coherence and increase $\rho_{xx}$ as the temperature is raised above $T_{min}$. In view of the scaling relation $\rho_{xy}^{AH}$ = $f(\rho_{xx})$, QCs are expected to contribute to $\rho_{xy}^{AH}(T)$ as well. The issue of whether or not QCs contribute to $\rho_{xy}^{AH}(T)$ has been addressed theoretically \cite{LW1991,Dug2001,MW2007}. These theories make the following specific predictions. (I) If the quantum interference effects are present in $\rho_{xx}$, WL correction to $\rho_{xy}^{AH}$ is \textit{finite} in \textit{low-resistivity} ferromagnetic metals at low temperatures where \textit{skew scattering} is important. (II) In \textit{high-resistivity} metallic ferromagnets or doped semiconductors in which the \textit{side-jump} mechanism prevails, WL correction to $\rho_{xy}^{AH}$ \textit{vanishes} even when the WL effect contributes to $\rho_{xx}$ for T $\lesssim$ T$_{min}$. (III) EEI corrections to $\rho_{xy}^{AH}$ \textit{vanish} for both skew scattering and side-jump mechanisms even when the EEI corrections to $\rho_{xx}$ are \textit{finite}. While the predictions (I) and (III) are vindicated by recent experimental findings on two-dimensional (2D) Fe films \cite{Mitra2007} and 3D Co$_{2}$FeSi films \cite{BKH2017}, respectively, the validity of the prediction (II) has not been tested so far. A complete understanding of how disorder affects $\rho_{xy}^{AH}$ is thus lacking at present.

\begin{figure}[htbp]
\centering
\includegraphics[scale=1.0, trim = 0 30 0 0, clip, width=0.8\linewidth]{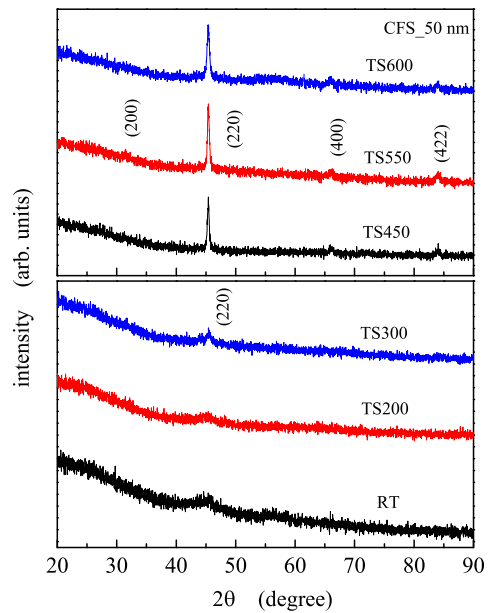}
\caption{Grazing incidence x-ray diffraction patterns of 50 nm thick Co$_{2}$FeSi thin film deposited at different substrate temperatures.}
\label{fig.1}
\vspace{-0.5cm}
\end{figure}

\begin{figure}[htbp]
\centering
\includegraphics[scale=1.0, trim = 0 30 0 60, clip, width=\linewidth]{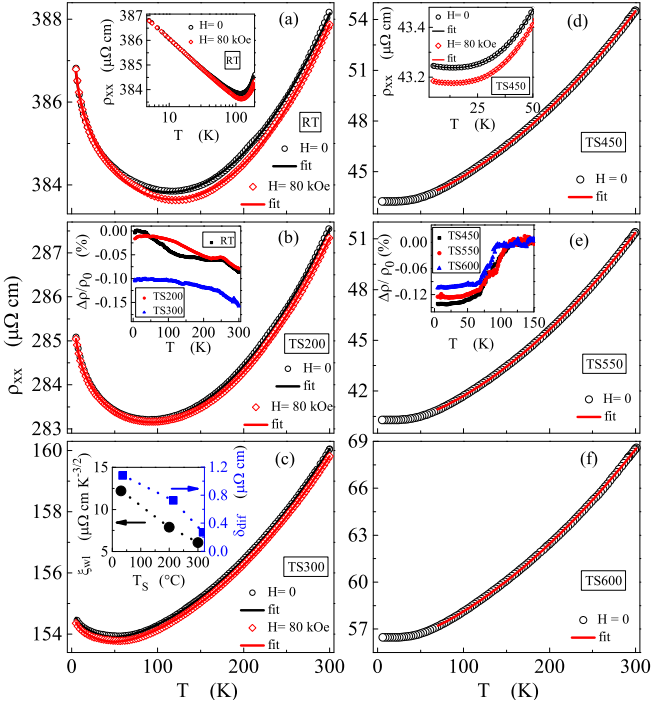}
\caption{Panels (a) - (c): $\rho_{xx}(T,H=0)$ and $\rho_{xx}(T,H=80kOe)$ (open circles) with the theoretical fits (continuous lines), based on Eq.(2). Panels (d) - (f): $\rho_{xx}(T,H=0)$ (open circles) along with the theoretical fits (continuous lines) in the range 70-300 K, based on Eq.(3) with $\gamma_{EEI}$ = 0. Inset of (a): the $-lnT$ dependence of $\rho_{xx}(H=0)$ and $\rho_{xx}(H=80kOe)$ below $T_{min}$ for the RT film. Inset of (d): $\rho_{xx}(T,H=0)$ and $\rho_{xx}(T,H=80kOe)$ along with the fits, based on Eq.(3), for the TS450 film over the temperature range 5-50 K. Insets of (b) and (e): magnetoresistance as a function of temperature for the amorphous and crystalline films, respectively. Inset of (c): the coefficients $\delta_{dif}$ and $\xi_{wl}$, in Eq.(2), versus substrate temperature, $T_{S}$, for the amorphous films.}
\label{fig.2}
\vspace{-0.5cm}
\end{figure}

\section{\label{sec:level2} Experimental details}
   
Co$_{2}$FeSi (CFS) Heusler-alloy thin films of 50 nm thickness were grown on the Si (111) substrate by ultra high vacuum dc magnetron sputtering at different substrate temperatures (T$_{S}$) ranging from room temperature (RT) to 600$^{\circ}$C. The details of the film deposition conditions and parameters can be found elsewhere \cite{BKH2017}. Figure 1 shows the grazing-incidence (grazing angle = 0.5$^{\circ}$) x-ray diffraction patterns, recorded on D8 Discover Br\"{u}ker x-ray diffractometer (Cu K$_{\alpha}$ source),  confirmed the amorphous nature of the CFS films deposited at room temperature (RT) and 200$^{\circ}$C whereas the film deposited at 300$^{\circ}$C is partially crystalline, indicated by the first appearance of a sharper (220) Bragg peak. The remaining films, grown at 450$^{\circ}$C, 550$^{\circ}$C and 600$^{\circ}$C, are in the fully-developed crystalline state. Based on the value of T$_{S}$, the CFS films are labeled as RT, TS200, TS300, TS450, TS550 and TS600. `Zero-field' electrical resistivity, $\rho_{xx}(T,H=0)$, and `in-field' resistivity, $\rho_{xx}(T,H=80 kOe)$, were measured using a four-probe setup equipped with a 80 kOe superconducting magnet (Oxford instruments). Hall resistivity was measured using AC transport option (five-probe setup) of the physical property measurement system of Quantum Design make. A wide variation in the degree of disorder, resulting in an order of magnitude variation in $\rho_{xx}$ of the CFS thin films, achieved by growing these films at different T$_{S}$, enables a detailed and systematic study of the effect of disorder on $\rho_{xy}^{AH}$ and $R_{A}$.

\section{\label{sec:level2} `Zero-field' and `in-field' electrical resistivity}

Figure 2 demonstrates that for all the films, $\rho_{xx}(T,H=0)$, and $\rho_{xx}(T,H=80 kOe)$ go through a minimum at a temperature, T$_{min}$, which decreases from $\simeq$ 115 K in the RT film to 12 K in the TS550 film. With increasing T$_{S}$, the residual resistivity, $\rho_{xx0}$ $\equiv$ $\rho_{xx}(T=5K,H=0)$, systematically decreases from 387 $\mu\Omega$ cm in the RT film to 40 $\mu\Omega$ cm in TS550. An elaborate analysis \cite{Babu1999,Srini1999,BKH2017} establishes that $\rho_{xx}(T,H=0)$ and $\rho_{xx}(T,H=80 kOe)$ are well described by the expression  

\begin{multline}
\rho_{xx}(T,H)= \rho_{xx0} - \delta_{dif}~lnT -\xi_{wl}~T^{3/2} + \beta_{e-m} ~ T^{2} \\ 
 + \alpha_{e-p} \int_0^{\theta_{D}/T}\frac{x^5}{(e^{x}-1)(1-e^{-x})}dx  
\end{multline}    

over the temperature range 5 K $\leq$ T $\leq$ 300 K in the amorphous films RT and TS200 (the partially crystalline film TS300) when $\alpha_{e-p} = 0$ ($\alpha_{e-p} \neq 0$ and $\theta_{D}$ is fixed at \cite{Bombor2013} 332 K). The second, third, fourth and fifth terms in Eq.(2) represent the contributions from the electron-diffuson ($e-d$) scattering \cite{Rivier1978,SNK1999}, weak localization \cite{LeeTVR1985,Babu1999,Srini1999} (WL), interband spin-flip electron-magnon ($e-m$) scattering and intraband non-spin-flip electron-phonon ($e-p$) scattering, respectively. The optimum fits, based on Eq.(2), are denoted by the continuous curves through the data (open circles) and displayed in Fig.2(a)-(c). Note that the $- lnT$ dependence of $\rho_{xx}(H=0)$ at low temperatures has been previously reported \cite{KKR1986,KKR1987} in a number of amorphous ferromagnetic alloys. 

By contrast, in the crystalline films TS450, TS550 and TS600, $\rho_{xx}(T,H=0)$ and $\rho_{xx}(T,H=80 kOe)$ are best described by the relation

\begin{multline}
\rho_{xx}(T,H)= \rho_{xx0} - \gamma_{EEI} ~ T^{1/2} + \beta_{e-m} ~ T^{2} \\ + \alpha_{e-p} \int_0^{\theta_{D}/T}\frac{x^5}{(e^{x}-1)(1-e^{-x})}dx 
\end{multline}    

in the temperature range 5 K $\leq$ T $\leq$ 70 K (70 K $\leq$ T $\leq$ 300 K), when the second term in Eq.(3), arising from the enhanced electron-electron Coulomb interaction, EEI, is finite (zero). The optimum fits (red continuous curves) to $\rho_{xx}(T,H=0)$ over the temperature range 70 K $\leq$ T $\leq$ 300 K, based on Eq.(3) with $\gamma_{EEI} = 0$, are shown in Fig.2(d)-(f) for the TS450, TS550 and TS600 films. A representative theoretical fit to $\rho_{xx}(T,H=0)$ and $\rho_{xx}(T,H=80 kOe)$ in the temperature range 5 K $\leq$ T $\leq$ 70 K, based on Eq.(3) with $\gamma_{EEI} \neq 0$, is displayed in the inset of Fig.2(d).

Note that the $-lnT$ variation of $\rho_{xx}$ at low temperatures in amorphous systems, resulting from the scattering of conduction electrons from the diffusive hydrodynamic modes associated with the longitudinal component of magnetization, called `diffusons' \cite{Rivier1978,SNK1999}, has been theoretically predicted \cite{Rivier1978} long ago. Since diffusons are \textit{insensitive} \cite{SNK1999} to $H$, the coefficient $\delta_{dif}$ of the $-lnT$ term in Eq.(2) is expected to have, at best, an extremely weak dependence on $H$. In conformity with this expectation, the inset of Fig.2(a) demonstrates that the $-lnT$ variation holds for both $\rho_{xx}(H=0)$ and $\rho_{xx}(H=80 kOe)$ for T $\lesssim$ 40 K and the magnetic field hardly affects this term (i.e., practically no change in the slope, $\delta_{dif}$, of the linear $\rho_{xx}(T,H)$ versus $lnT$ plot). Insensitivity of $\delta_{dif}$ to $H$ completely rules out the Kondo mechanism for the $-lnT$ term. Another possible origin of the $-lnT$ term could be the EEI and/or WL QC in 2D systems \cite{LeeTVR1985} but this possibility is highly unlikely because a 50 nm thick CFS film cannot be regarded as a 2D system. It immediately follows that, (i) for T $\lesssim$ 40 K, the $e-d$ scattering almost entirely accounts for $\rho_{xx}(T)$ and for negligibly small magnetoresistance, MR, ($\Delta\rho/\rho_{0}$, evident from the inset of Fig.2(b)) in the amorphous films RT and TS200. However, in these films, WL and $e-m$ contributions dominate over the $e-d$ one at higher temperatures and lead to negative MR \cite{LeeTVR1985,SNK2005}. By comparison, the $e-d$ scattering contribution is relatively less important in the partially crystallized film TS300 in which WL and $e-m$ scattering cause negative MR. (ii) Along with the $e-d$ scattering, WL and $e-m$ (as well as $e-p$) contributions are required to reproduce the observed $\rho_{xx}(T,H)$ in the entire temperature range 5 K - 300 K for the RT and TS200 films (TS300 film). Absence of the $e-p$ contribution of the Bloch-Gr\"{u}neisen (BG) form (given by the last term in Eq.(2)) in the amorphous films RT and TS200 can be understood as follows. The modified diffraction model, more appropriate \cite{KKR1986,KKR1987} for the amorphous systems than the BG model, considers the scattering of conduction electrons from the potential of the disordered spatial arrangement of atoms and predicts that $\rho_{xx}$ $\sim$ $T^{2}$ for $T < \theta_{D}$. Thus, besides a dominant $e-m$ scattering contribution, the $T^{2}$ term in Eq.(2) can have a significant contribution due to the scattering of conduction electrons from the structural-disorder, particularly for the CFS amorphous films RT and TS200. (iii) As a consequence of the diminished amorphous phase (volume) fraction as T$_{S}$ approaches $300^{\circ}$C, $\delta_{dif}$ as well as $\xi_{wl}$ are considerably reduced (inset of Fig.2(c)) so much so that they are insignificant in the crystalline CFS films TS450, TS550 and TS600. Instead, in these films, the $e-d$ and WL contributions completely absent and  EEI QC solely determines $\rho_{xx} (T)$ (Cf. Eq.(2) and Eq.(3)) while negative MR (inset of Fig.2(e)) essentially results from the suppression \cite{BKH2017} of the $e-m$ scattering by $H$. (iv) The $e-p$ contribution has hardly any variation (within the uncertainty limits) with T$_{S}$ in the range $450^{\circ}$C $\leq T_{S} \leq$ $600^{\circ}$C because the phonon spectrum is the same in the crystalline CFS films TS450, TS550 and TS600. In sharp contrast, the coefficient $\beta_{e-m}$ initially falls rapidly as T$_{S}$ increases towards $300^{\circ}$C and thereafter the rate of decline slows down, as is evident from the Fig.3(d). This variation of $\beta_{e-m}$ with $T_{S}$ finds a straightforward explanation in terms of the prediction \cite{BKH2017,SNK2005} $\beta_{e-m} \sim  M_{s}(0) ~ [D(T)]^{-2}$ with \cite{SNK1999} $D(T) = D(0) ~ (1 - D_{2}T^{2} - D_{5/2}T^{5/2})$, where $D$ is the spin-wave stiffness, $D(0)$ and $M_{s}(0)$ are $D$ and spontaneous magnetization at 0 K, $D_{2}$ and $D_{5/2}$ account for the thermal renormalization of $D$ due to Stoner single particle-magnon and magnon-magnon interactions. The quantities $M_{s}(0)$ and $D(0)$ are determined from $M_{s}(T)$, as elucidated in the next section.

\section{\label{sec:level2} Spontaneous magnetization and spin waves}

Fig.3(a) depicts the $M_{\parallel}(H)$ isotherms taken on the TS550 Co$_{2}$FeSi (CFS) thin film at temperatures in the range 5 K - 300 K when the magnetic field, $H$, is applied within the film plane. These isotherms are typical of other CFS films as well. As illustrated in this figure, spontaneous magnetization at different temperatures, $M_{s}(T)$, is computed from the intercepts on the ordinate obtained when the linear high-field portions of the $M_{\parallel}(H)$ isotherms are extrapolated to $H = 0$. $M_{s}(T)$ data (symbols), so obtained, are shown in Fig.3(b). The least-squares fits (continuous curves through the $M_{s}(T)$ data) are attempted based on the well-known spin-wave (SW) expression\cite{SNK1999,SNK1983,SNK1991,KB1994,KB1998} 
\begin{multline}
M_{s}(T) = M_{s}(0) - g ~ \mu_{B} ~ \\
\times\left[ \zeta(3/2)  \left( \frac{k_{B} T}{4 \pi D(T)} \right)^{3/2} +
15 \pi ~ \beta ~ \zeta(5/2)  \left (\frac{k_{B} T}{4 \pi D(T)} \right)^{5/2} \right]  
\end{multline}    
with $\beta$ = $\langle r^{2}\rangle$/20, where $\langle r^{2} \rangle$ is the mean-square range of the exchange interaction. In Eq.(4), $M_{s}(0)$, $D_{0}$, $D_{2}$, $D_{5/2}$ and $\beta$ are varied so as to obtain the best SW fits. This exercise reveals that $D_{2}$, $D_{5/2}$ and $\beta$ are inconsequential. That $D_{2}$ = $D_{5/2}$ = $\beta$ $\cong$ 0 is clearly borne out by the linear $M_{s}$ versus T$^{3/2}$ plots, shown in Fig.3(b). The optimal SW fits (the continuous straight lines) to the $M_{s}(T)$ data and the corresponding values of the parameters $M_{s}(0)$ and $D(0)$, as functions of T$_{S}$, for different CFS thin films are displayed in Fig.3(b) and 3(c), respectively. Clearly, the observed decreasing trend of $\beta_{e-m}$ with T$_{S}$ is qualitatively reproduced (Fig.3(d)) when the presently determined values of $M_{s}(0)$ and $D(0)$ for different $T_{S}$ are used in the relation \cite{BKH2017,SNK2005} $\beta_{e-m} \sim  M_{s}(0) ~ [D(0)]^{-2}$ for $\beta_{e-m}$.

Considerably reduced $D(0)$ for the amorphous films RT and TS200 (Fig.3(c)) compared to that for the crystalline films TS450, TS550 and TS600 finds the following interpretation in terms of the spin-fluctuation model \cite{SNK1999}. In disordered (amorphous) ferromagnets, magnons and diffusons both contribute \cite{SNK1999} to $M_{s}(T)$ and both give rise to the T$^{3/2}$ variation \cite{SNK1999} of $M_{s}$. Note that the slope of the linear $M_{s}$ - T$^{3/2}$ plots is $\propto$ $D(0)^{-3/2}$. \textit{Increased slope} (or equivalently, reduced $D(0)$) in the amorphous films RT and TS200 compared to other films in Fig.3(b), basically reflects a sizable diffuson contribution (apart from that due to magnons) to $M_{s}(T)$, as is the case for $\rho_{xx}(T)$.

\begin{figure}[htbp]
\centering
\includegraphics[scale=1.0, trim = 0 10 0 10, clip, width=\linewidth]{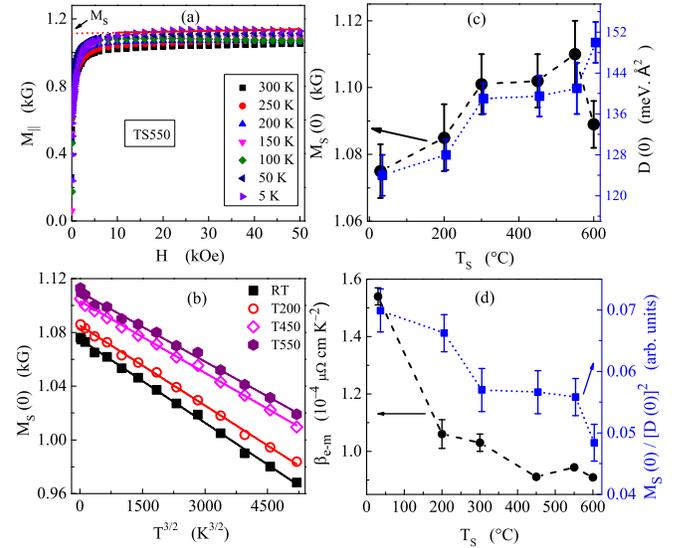}
\caption{Panel (a): Magnetization ($M$) versus magnetic field ($H$) isotherms at different temperatures (5-300 K) for the TS550 film. These isotherms are representative of other films as well. Panel (b): $M_{s}$ versus T$^{3/2}$ plots. Panel (c): $M_{s}(0)$ and $D(0)$ as functions of T$_{S}$. Panel (d): Variations with the substrate temperature, $T_{S}$, of the coefficient ($\beta_{e-m}$) of the electron-magnon scattering contribution to `zero-field' resistivity and the ratio, $M_{S}(0)$ / $[D(0)]^{2}$, for the Co$_{2}$FeSi thin films.}
\label{fig.3}
\vspace{-0.5cm}
\end{figure}

\section{\label{sec:level2} Anomalous Hall effect}

Insets of Fig.4(a)-(d) display $\rho_{xy}^{AH}(T)$ for the RT, TS200, TS300 and TS450 Co$_2$FeSi thin films, obtained by extrapolating the linear high-field portions of the Hall resistivity ($\rho_{xy}$) versus $H$ isotherms\cite{BKH2017} to $H$ = 0. At any given temperature, $\rho_{xy}^{AH}$ systematically decreases as the crystalline fraction increases. $\rho_{xy}^{AH}(T)$, like $\rho_{xx}(T)$, exhibits a minimum at T$_{min}$ $<$ 100 K (T$_{min}$ in $\rho_{xy}^{AH}$ is lower than that in $\rho_{xx}$) in the \textit{disordered} films RT, TS200 and TS300 whereas such a minimum, though \textit{present} in $\rho_{xx}(T)$, is \textit{completely absent} in $\rho_{xy}^{AH}(T)$ in the \textit{crystalline}, `ordered', films TS450, TS550 and TS600. In this context, recall that, in the disordered films, the $e-d$ and WL terms in Eq.(2), i.e., $-\delta_{dif}~lnT$ and $-\xi_{wl}~T^{3/2}$, compete with the $e-m$ term, $\beta_{e-m} ~ T^{2}$, to produce the resistivity minimum. On the other hand, the best fits (red curves in the insets of Fig.4(a) - (c)) to the $\rho_{xy}^{AH}(T)$ data (symbols) for T$\leq$150 K assert that only the terms $-\delta_{dif}~lnT$ and $\beta_{e-m} ~ T^{2}$ entirely account for the minimum in $\rho_{xy}^{AH}(T)$ and the WL QC is of no consequence. This is also true for $R_{A}(T)$ = $\rho_{xy}^{AH}(T)$ / $4\pi~M_{s}(T)$, in which the minima are shifted to yet lower temperatures. In sharp contrast, the EEI QC, basically responsible for the upturn in $\rho_{xx}(T)$ at T $\lesssim$ $T_{min}$ in the crystalline films, does not contribute to $\rho_{xy}^{AH}(T)$ or $R_{A}(T)$ with the result that these quantities do not exhibit a minimum. Since all the mechanisms contributing to $\rho_{xx}(T)$ do not get reflected in $\rho_{xy}^{AH}(T)$ and the magnitudes of even the \textit{common} contributions are different, the scaling relation, $\rho_{xy}^{AH}$ = $f(\rho_{xx})$ breaks down completely (Fig.5(a)) for all the films, more so in the disordered ones. Thus, the correct form of scaling can be arrived at by subtracting the $e-d$ contribution ($e-d$ and WL contributions) from the raw $R_{A}(T)$ ($\rho_{xx}(T)$) data for the disordered films RT, TS200 and TS300, and by subtracting the EEI contribution from the raw $\rho_{xx}(T)$ data for the crystalline films. The $R_{A}(T)$ and $\rho_{xx}(T)$ data, so corrected, are shown in figures 4 and 6, respectively, along with the corresponding raw data.

Assuming the validity of Matthiessen's rule, longitudinal resistivity can be written as $\rho_{xx}$ = $\rho_{xx0}$ + $\rho_{xxT}$, where $\rho_{xx0}$ is the residual (temperature-independent) resistivity and $\rho_{xxT}$ is the intrinsic (temperature-dependent) resistivity. Substituting this expression for $\rho_{xx}$ in Eq.(1) yields

\begin{align}
R_{A}(T)&=(a  \rho_{xx0}+a  \rho_{xxT})+(b  \rho_{xx0}^2+2b ~ \rho_{xx0} ~ \rho_{xxT} + b  \rho_{xxT}^2)
\end{align}

\begin{figure}[htbp]
\centering
\includegraphics[scale=1.0, trim = 0 0 0 10, clip, width=\linewidth]{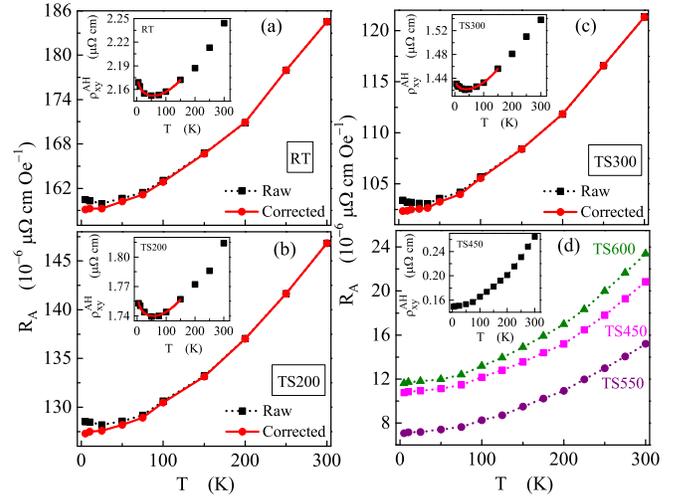}
\caption{(a)-(d): \textit{Raw} and \textit{corrected} (\textit{raw}) R$_{A}(T)$ for the RT, TS200, TS300 (TS450, TS500, TS600) films. The insets: \textit{raw} $\rho_{xy}^{AH}(T)$ data for the RT, TS200, TS300 and TS450 films.}
\label{fig.4}
\vspace{-0.5cm}
\end{figure}

\begin{figure}[htbp]
\centering
\includegraphics[scale=1.0, trim = 0 0 0 220, clip, width=\linewidth]{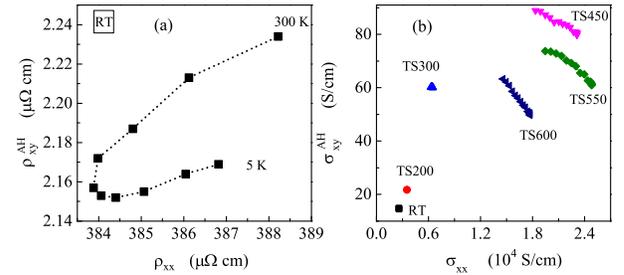}
\caption{(a) Anomalous Hall resistivity, $\rho_{xy}^{AH}$, versus longitudinal resistivity, $\rho_{xx}$. (b) Anomalous Hall conductivity, $\sigma_{xy}^{AH}$, versus longitudinal conductivity, $\sigma_{xx}$.}
\label{fig.5}
\vspace{-0.5cm}
\end{figure}

For a given film, the best theoretical fit to the \textit{corrected} $R_{A}(T)$, based on Eq.(5), is obtained by optimizing the coefficients $a$ and $b$, and using the values of $\rho_{xx0}$ and $\rho_{xxT}$ from the \textit{corrected} $\rho_{xx}(T)$ data. The panels (a) and (b) of figure 7 highlight the temperature variation of $R_{A}$ (continuous curves) and that of each individual term in Eq.(5), yielded by the best fit based on Eq.(5), for the films RT and TS550. These temperature variations are representative of other amorphous and crystalline thin films as well. From Fig.7(a),(b), it is obvious that the cross-term (2$b~\rho_{xx0}$ $\rho_{xxT}$) essentially governs the temperature dependence of $R_{A}$ in all the films since the term, $b ~ \rho_{xxT}^2$, (originating from Sj or intrinsic KL/Bc or both) and the Sk term, $a ~ \rho_{xxT}$, have \textit{weak} but \textit{competing} variations with temperature. As T $\rightarrow$ 0 K, sizable (but opposite in sign) contributions to $R_{A}$ come from the Sj ($b ~ \rho_{xx0}^2$) and Sk ($a ~ \rho_{xx0}$). These contributions are much larger in the amorphous than in crystalline CFS films because of the order of magnitude larger $\rho_{xx0}$ in the former case.

\begin{figure}[htbp]
\centering
\includegraphics[scale=1.0, trim = 0 130 0 140, clip, width=\linewidth]{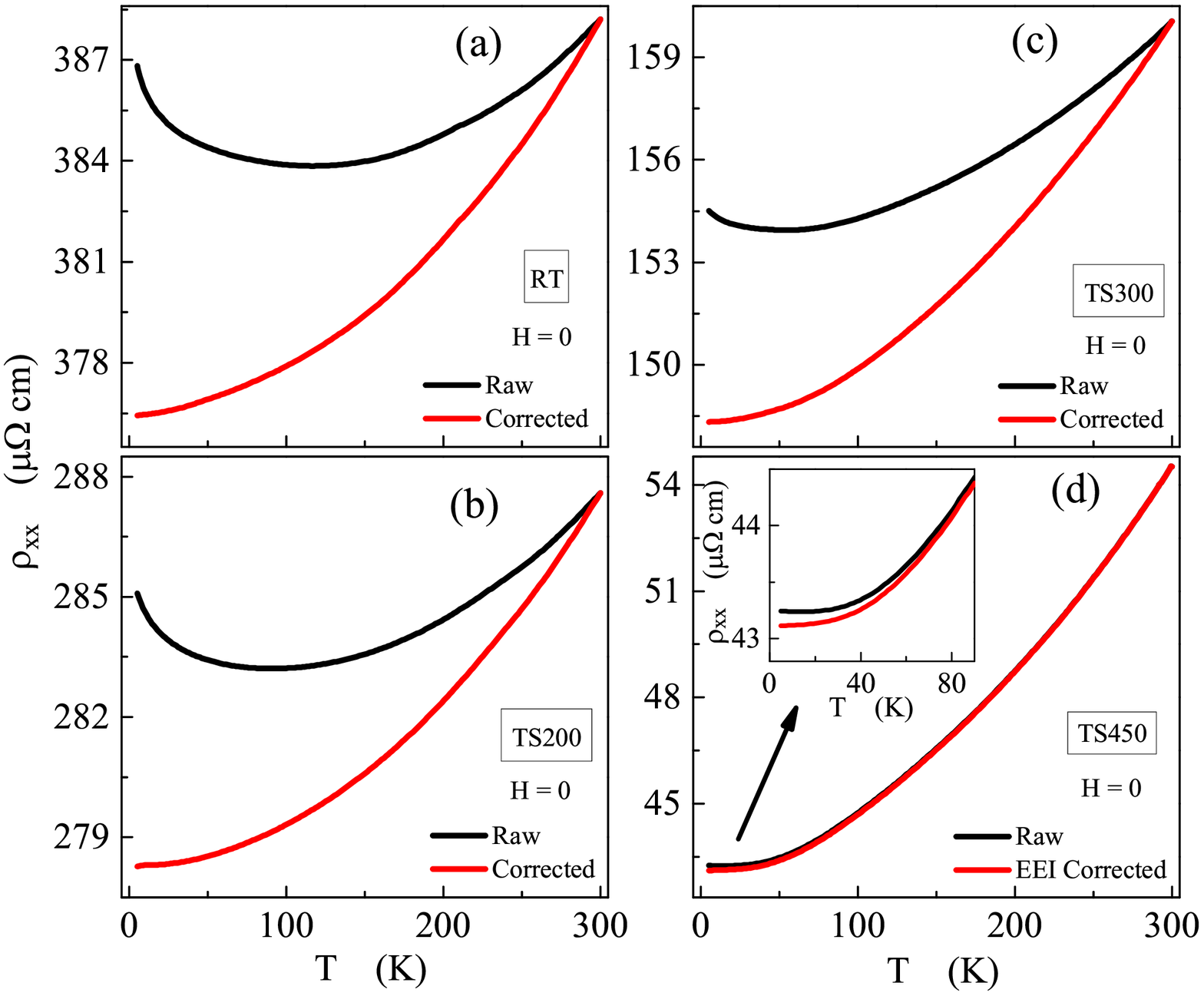}
\caption{(a)-(d): Raw (black curves) and corrected (red curves) $\rho_{xx}(T)$ data for the RT, TS200, TS300 and TS450 (representative of TS550 and TS600 as well) films. Inset of (d): magnified view of the low-temperature raw and corrected $\rho_{xx}(T)$ data for the TS450 film.}
\label{fig.6}
\vspace{0.0cm}
\end{figure}

In order to bring out explicitly the direct relation between $R_{A}(T)$ and $\rho_{xxT}$, Eq.(5) can be rewritten in the form

\begin{equation}
[R_{A}(T) - R_{A}(0)] = (a + 2b ~ \rho_{xx0}) ~ \rho_{xxT} + b ~ \rho_{xxT}^2
\end{equation} 

where $R_{A}(0)= a ~ \rho_{xx0} + b ~ \rho_{xx0}^2$. A nearly linear variation of $[R_{A}(T) - R_{A}(0)]$ with $\rho_{xxT}$, evident in Fig.7(c),(d), suggests that the dominating contribution comes from the cross-term, which is largely due to the Sj mechanism. The scaling relation, Eq.(6), holds for all the 50 nm thick CFS films regardless of the degree of disorder present, and even for the well-ordered CFS thin films of different thicknesses \cite{BKH2017}. From this result, we conclude that the side-jump mechanism almost entirely accounts for the anomalous Hall effect in ordered as well as disordered Co$_2$FeSi Heusler-alloy thin films, and that the correct form of scaling is $[R_{A}(T) - R_{A}(0)]$ = $f(\rho_{xxT})$, i.e., Eq.(6), and not $\rho_{xy}^{AH}$ = $f(\rho_{xx})$, which is found to break down (Fig.5(a)) in all the CFS films \cite{BKH2017}. Consistent with this finding, the customary approach of directly relating anomalous Hall conductivity (AHC), $\sigma_{xy}^{AH}$, to longitudinal conductivity, $\sigma_{xx}$, also fails in the present case (Fig.5(b)). In the clean region \cite{Onoda2006,Miya2007} ($10^4\leq \sigma_{xx} \leq10^6 S/cm$), the theory predicts \cite{Miya2007} the value $\sigma_{xy}^{AH} \simeq$ 10$^{3}$ S/cm if the intrinsic mechanism solely determines AHC. Though $\sigma_{xx}$ falls within the clean regime in the crystalline CFS films, $\sigma_{xy}^{AH}$ is more than one order of magnitude smaller (Fig.5(b)). A strongly suppressed AHC basically reflects dominant side-jump mechanism \cite{BKH2017}.

\begin{figure}[htbp]
\centering
\includegraphics[scale=1.0, trim = 0 0 0 10, clip, width=\linewidth]{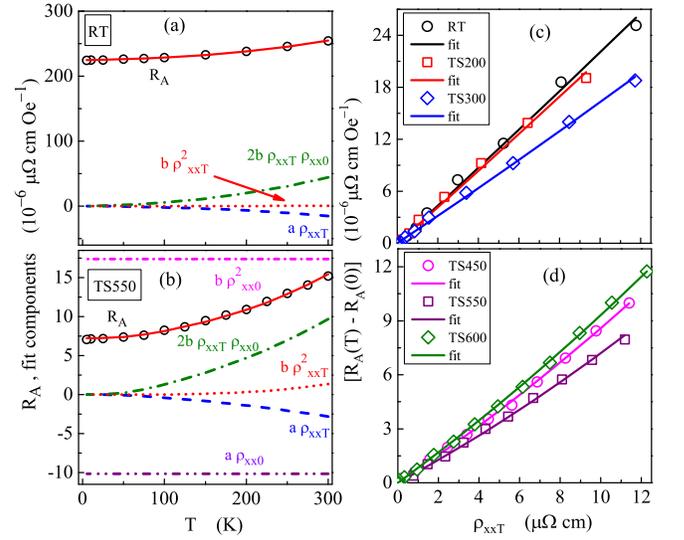}
\caption{(a),(b): Temperature variations of the individual terms in Eq.(5) along with the theoretical fit (continuous curves) to R$_{A}(T)$ (open circles), based on Eq.(5), for the RT and TS550 films. (c),(d): Scaling of [$R_{A}(T) - R_{A}(0)$] (open symbols) with $\rho_{xxT}$; the solid lines represent the theoretical fits based on Eq.(6). }
\label{fig.7}
\vspace{-0.5cm}
\end{figure}

\section{\label{sec:level6} Summary and Conclusion} 

From an elaborate analysis of the `zero-field' electrical resistivity, $\rho_{xx}(T)$, `in-field' resistivity, $\rho_{xx}(T,H = 80 kOe)$, magnetoresistance (MR) and anomalous Hall resistivity, $\rho_{xy}^{AH}(T)$, and magnetization , $M$, of the Co$_2$FeSi (CFS) Huesler alloy thin films of fixed (50 nm) thickness, prepared in different states of disorder, permits us to draw the following conclusions. Of all the mechanisms contributing to anomalous Hall effect (AHE), side-jump essentially determines $\rho_{xy}^{AH}$ or $R_{A}$ in the CFS films, regardless of the degree of disorder present. In the (high-resistivity) amorphous films RT, TS200 and TS300, the electron-diffuson, $e-d$, scattering and weak localization (WL) mechanisms both compete with the inelastic electron-magnon, $e-m$, scattering to give rise to the minimum in $\rho_{xx}(T)$; by comparison, the minimum in $R_{A}(T)$ or $\rho_{xy}^{AH}(T)$ originates from the competing $e-d$ and $e-m$ mechanisms, as WL does not contribute to AHE. In sharp contrast, in the TS450, TS550 and TS600 (low-resistivity) crystalline films, the enhanced electron-electron Coulomb interaction (EEI), which causes the upturn in resistivity at temperatures T$<$T$_{min}$, makes no contribution to $\rho_{xy}^{AH}(T)$, with the result that AHR does not exhibit a minimum. The customary practice of scaling $\rho_{xy}^{AH}$ with $\rho_{xx}$ or $\sigma_{xy}^{AH}$ with $\sigma_{xx}$ beaks down completely in the present case. Instead, when $\rho_{xy}^{AH}(T)$ is corrected for the $e-d$ contribution and $\rho_{xx}(T)$ for both $e-d$ and WL contributions (only EEI) in the amorphous (crystalline) films, $R_{A}(T)$, calculated from the \textit{corrected} $\rho_{xy}^{AH}$, perfectly scales with $\rho_{xxT}$, the temperature-dependent part of the \textit{corrected} $\rho_{xx}$, for all the CFS thin films. 

This work vindicates the long-standing theoretical prediction \cite{LW1991,Dug2001,MW2007} that in \textit{high-resistivity} metallic ferromagnets in which the \textit{side-jump} mechanism prevails, WL correction to $\rho_{xy}^{AH}$ \textit{vanishes} even when the WL effect contributes to $\rho_{xx}$ for T $\lesssim$ T$_{min}$, and provides a strong motivation for undertaking the theoretical calculations that address the (hitherto unexplored) role of diffusons in anomalous Hall effect in disordered ferromagnets.

\section*{\label{sec:level2} Acknowledgement}
B. K. Hazra acknowledges the financial assistance (SRF) from UGC-BSR and thanks the Centre for Nanotechnology, University of Hyderabad, for allowing the use of the Physical Property Measurement System. This work was supported by Indian National Science Academy under Grant No.: SP/SS/2013/1971.


\begin{thebibliography}{50}    


\bibitem{Smit1955} J. Smit, {\it Physica} {\bf 21}, 877 (1955); J. Smit, {\it Physica} {\bf 24}, 39 (1958).
\bibitem{Berger1970} L. Berger, {\it Phys. Rev.} B {\bf 2}, 4559 (1970). 
\bibitem{KL1954} R. Karplus and J. M. Luttinger, {\it Phys. Rev.} {\bf 95}, 1154 (1954); J. M. Luttinger, {\it Phys. Rev.} {\bf 112}, 739 (1958). 
\bibitem{Yao2004} Y. Tao, L. Kleinman, A. H. MacDonald, J. Sinova, T. Jungwirth, D. S. Wang, E. Wang and Q. Niu, {\it Phys. Rev. Lett.}  {\bf 92}, 037204 (2004).
\bibitem{Naga2010} N. Nagaosa, J. Sinova, S. Onoda, A. H. MacDonald and N. P. Ong, {\it Rev. Mod. Phys.} {\bf 82}, 1539 (2010) and references cited therein.
\bibitem{Xiao2010} D. Xiao, C. Chang and Q. Niu, {\it Rev. Mod. Phys.} {\bf 82}, 1959 (2010) and references cited therein.
\bibitem{Kuebler2012} J. K\"{u}bler and C. Felser, {\it Phys. Rev.} B {\bf 85}, 012405 (2012).
\bibitem{Kaul1979} S. N. Kaul, {\it Phys. Rev.} B {\bf 20}, 5122 (1979); S. N. Kaul, {\it Phys. Rev.} B {\bf 15}, 7552 (1977).
\bibitem{Onoda2006} S. Onoda, N. Sugimoto and N. Nagaosa, {\it Phys. Rev. Lett.}  {\bf 97}, 126602 (2006).
\bibitem{Hou2015} D. Hou, G. Su, Y. Tian, X. Jin, S. A. Yang and Q. Niu {\it Phys. Rev. Lett.}  {\bf 114}, 217203 (2015).
\bibitem{Hyodo2016} K. Hyodo, A. Sakuma and Y. Kota, {\it Phys. Rev.} B {\bf 94}, 104404 (2016).
\bibitem{Miya2007} T. Miyasato, N. Abe, T. Fujii, A. Asamitsu, S. Onoda,Y. Onose, N. Nagaosa, and Y. Tokura, {\it Phys. Rev. Lett.}  {\bf 99}, 086602 (2007).
\bibitem{Tian2009} Y. Tian, L. Ye and X. Jin, {\it Phys. Rev. Lett.}  {\bf 103}, 087206 (2009).
\bibitem{Xiong2010} Y. M. Xiong, P. W. Adams and G. Catelani, {\it Phys. Rev. Lett.}  {\bf 104}, 076806 (2010).
\bibitem{Vidal2011} E. V. Vidal, H. Schneider and G. Jakob, {\it Phys. Rev.} B {\bf 83}, 174410 (2011).
\bibitem{Ye2012} L. Ye, Y. Tian, X. Jin and D. Xiao, {\it Phys. Rev.} B {\bf 85}, 220403 (2012).
\bibitem{Presti2014} J. C. Prestigiacomo, D. P. Young, P. W. Adams and S. Stadler, {\it J. Appl. Phys.} {\bf 115}, 043712 (2014).
\bibitem{Zhu2015} T. Zhu and S. B. Wu, {\it IEEE Trans. Magn,} {\bf 51}, 4400604 (2015). 
\bibitem{Obaida2011} M. Obaida, K. Westerholt, and H. Zabel, {\it Phys. Rev.} B {\bf 84}, 184416 (2011).
\bibitem{Zhu2012} Zhu Qin, Xin-Dian Liu, and Zhi-Qing Li, {\it J. Appl. Phys.} {\bf 111}, 083919 (2012).
\bibitem{Gabor2015} M. S. Gabor, M. Belmeguenai, T. Petrisor, Jr., C. Ulhaq-Bouillet, S. Colis, and C. Tiusan, {\it Phys. Rev.} B {\bf 92}, 054433 (2015).
\bibitem{Lin2016} L. Wu, K. Zhu, D. Yue, Y. Tian, and X. Jin, {\it Phys. Rev.} B {\bf 93}, 214418 (2016).
\bibitem{Guo2012} Z. B. Guoa, W. B. Mi, Q. Zhang, B. Zhang, R. O. Aboljadayel and X. X. Zhan, {\it Solid State Comm.} {\bf 152}, 220-224 (2012).
\bibitem{Lu2013} Y. M. Lu, J. W. Cai, Z. Guo and X. X. Zhang, {\it Phys. Rev.} B {\bf 87}, 094405 (2013).
\bibitem{LeeTVR1985} P. A. Lee and T. V. Ramakrishnan, {\it Rev. Mod. Phys.} {\bf 57}, 287 (1985).
\bibitem{Babu1999} P. D. Babu, S. N. Kaul, L. Fernand\'{e}z Barqu\'{i}n, J. C. Gom\'{e}z Sal, W. H. Kettler and M. Rosenberg, {\it Int. J. Mod. Phys.} B {\bf 13}, 141 (1999).
\bibitem{Srini1999} S. Srinivas, S. N. Kaul and S. N. Kane, {\it J. Non-cryst. Solids} {\bf 248}, 211 (1999).
\bibitem{LW1991} A. Langenfeld and P. W\"{o}lfle {\it Phys. Rev. Lett.}  {\bf 67}, 739 (1991).
\bibitem{Dug2001} V. K. Dugaev, A. Cr\'{e}pieux and P. Bruno, {\it Phys. Rev.} B {\bf 64}, 104411 (2001).
\bibitem{MW2007} K. A. Muttalib and P. W\"{o}lfle, {\it Phys. Rev.} B {\bf 76}, 214415 (2007).
\bibitem{Mitra2007} P. Mitra, R. Misra, A. F. Hebard, K. A. Muttalib and P. W\"{o}lfle, {\it Phys. Rev. Lett.} {\bf 99}, 046804 (2007).
\bibitem{BKH2017} B. K. Hazra, S. N. Kaul, S. Srinath, M. Manivel Raja, R. Rawat and Archana Lakhani, {\it Phys. Rev.} B {\bf 96}, 184434 (2017).
\bibitem{Bombor2013} D. Bambor, C. G. F. Blum, O. Volkonskiy, S. Rodan, S. Wurmehl, C. Hess and B. Buchner, {\it Phys. Rev. Lett.}  {\bf 110}, 066601 (2013).
\bibitem{Rivier1978} M A Continentino and N Rivier, {\it J. Phys. F: Metal Phys.} {\bf 8}, 1187 (1978).
\bibitem{SNK1999} S. N. Kaul, {\it J. Phys.: Condens. Matter} {\bf 11}, 7597 (1999); A. Semwal and S. N. Kaul, {\it Phys. Rev.} B {\bf 60}, 12799 (1999).  
\bibitem{KKR1986} S. N. Kaul, W. Kettler and M. Rosenberg, {\it Phys. Rev.} B {\bf 33}, 4987 (1986). 
\bibitem{KKR1987} S. N. Kaul, W. Kettler and M. Rosenberg, {\it Phys. Rev.} B {\bf 35}, 7153 (1987).
\bibitem{SNK2005} S. N. Kaul, {\it J. Phys.: Condens. Matter} {\bf 17}, 5595 (2005).
\bibitem{SNK1983} S. N. Kaul, {\it Phys. Rev.} B {\bf 27}, 5761 (1983); S. N. Kaul, {\it Phys. Rev.} B {\bf 27}, 6923 (1983).
\bibitem{SNK1991} S. N. Kaul, {\it J. Phys.: Condens. Matter} {\bf 3}, 4027 (1991).
\bibitem{KB1994} S. N. Kaul and P. D. Babu, {\it Phys. Rev.} B {\bf 50}, 9308 (1994).
\bibitem{KB1998} S. N. Kaul and P. D. Babu, {\it J. Phys.: Condens. Matter} {\bf 10}, 1563 (1998). 




\end{thebibliography}
\end{document}